# Quantum point contact with large localized spin: fractional quantization of the ballistic conductance

I.A. Shelykh[1,2], N.G. Galkin[3] and N.T. Bagraev[4]

[1] *Physics and Astronomy School, University of Southampton, Highfield, Southampton, SO17 1BJ, UK*

[2] *St.Petersburg State Polytechnical University, Polytechnicheskaya 29, 195251, St. Petersburg, Russia*

[3] *Algodign LLC, Force Field Lab, 123379, Moscow, Russia*

[4] *A.F.Ioffe Physico- Technical Institute of RAS, 194021, St. Petersburg, Russia*

We analyze the conductance of the quantum point contact containing large localized spin *J*. The additional plateau is formed on a ballistic conductance staircase if only one propagating channel is rendered conducting. The conductance value at this plateau is shown to depend strongly on *J* and decrease from $3e^2/2h$ to $e^2/h$ when *J* increases from ½ to infinity, which is in a good agreement with the experimental observations [D.J. Reilly, et al, Phys. Rev. B **63**, 121311 (2001)].

The progress in nanotechnology allowed the fabrication of the quasi- one dimensional mesoscopic components in which the transport of the carriers has a ballistic character and is not accompanied by the Joule losses. The conductance of such structures is determined by the number of the filled subbands of the dimensional quantization *N* and can be calculated using the Landauer- Buttiker formula [1]

$$G = g_s \frac{e^2}{h} \sum_{i,j=1}^{N} T_{ij} \qquad (1)$$

where e and h are the electron charge and the Plank constant, the spin factor $g_s = 2$ if the subbands are spin- degenerated, and $g_s = 1$ if this degeneracy is lifted, $T_{ij}$ are transition coefficients from i-th to j-th propagating mode, $\sum_{i,j=1}^{N} T_{ij} < 1$.

The role of the spin- correlations in the ballistic transport is sufficiently enhanced as compare to the classical transport in the diffusive regime, specifically if only one propagating channel is rendered conducting ($N=1$). Among their most drastic manifestations is the findings of the "$0.7 \cdot (2e^2/h)$" feature at the first step of the conductance staircase of the quantum wires (QWs) and quantum point contacts (QPCs) at zero magnetic field [2, 3]. It is now almost generally accepted that this anomaly is connected with an appearance of the spontaneously spin-polarized state in the region of the QPC. Two experimental observations support this conclusion. First, the electron $g$ - factor was found to increase from 0.4 to 1.3 as the number of occupied 1D subbands decreases [3]. Second, the height of the "$0.7 \cdot (2e^2/h)$" feature attains a value of $0.5 \cdot (2e^2/h)$ with increasing of the external magnetic field [2, 3]. Theoretical calculations carried out within the LDA [4] or the Hartree-Fock [5] approximations demonstrate the possibility of the formation of the spontaneously spin-polarized state in the low concentration limit, when the exchange interaction between electrons dominates over their kinetic energy. The role of the mutual orientation of the spins of localized and propagating electrons was also important to be discussed [6, 7]. If the external magnetic field is zero and only single electron is localized at the QPC, the conductance value was shown to be equal to $3/4 \cdot (2e^2/h)$ in the case of the ferromagnetic exchange interaction (when the energy lowers) or $1/4 \cdot (2e^2/h)$ in the case of the antiferromagnetic exchange interaction. The former value is in a good agreement with the experimental data.

Since this conclusion is of importance for our further consideration, here we discuss briefly how this result has been obtained as well as consider the conductance of the QPC containing the spin S>1/2. The physical model of this hypothetical situation seems to be the $Mn^{2+}$ ion with spin S=5/2 embedded in the QPC. Alternatively, it could be the QPC containing several localised electrons exhibiting the spontaneous spin-polarization due to the exchange interaction. The spin- dependent scattering of the propagating electrons in such a system can also manifest itself in the formation of the additional plateaux of the quantum conductance staircase. The conductance value is expected, however, to be different from $0.75 G_0$ ($G_0 = 2e^2/h$) and to be dependent on the number of the electrons with unpaired spin localized on the QPC.

Firstly, the situation can be considered qualitatively. The contact is supposed to contain total spin $J$. Having nonzero magnetic moment the localized state affects the propagating carriers via the Kondo- type correlations. As the interaction between the localized and propagating electrons depends on the mutual orientation of their spins, the transmission coefficient through the QPC with a magnetic moment becomes to be spin-dependent. If the propagating electron enters the QPC, the total spin $S$ can be either $S_1 = J + 1/2$ or $S_2 = J - 1/2$. The number of the possible realizations of the state 1 appears to be $N_1 = 2S_1 + 1 = 2J + 2$, while the number of the realizations of the state 2 is $N_2 = 2S_2 + 1 = 2J$, with the configurations 1 and 2 split in energy because of the exchange interaction. In the case of the ferromagnetic interaction the energy of the state 1 lies below the energy of the state 2, and thus the potential barrier formed in the region of the QPC is higher for the configuration 2. Consequently, for small enough carrier concentration the ingoing electron in the configuration 1 passes freely the QPC, while in the configuration 2 it is reflected. Then, the only configuration 1 contributes to the conductance. The probability of its realization is $(J+1)/(2J+1)$ in zero external magnetic field against the $J/(2J+1)$ for the configuration 2, and thus the conductance of the QPC in the considered regime reads

$$G_f = \frac{J+1}{2J+1} G_0 \qquad (2)$$

On the contrary, in the case of the antiferromagnetic interaction, the configuration 2 is energetically preferable, and the conductance should be

$$G_a = \frac{J}{2J+1} G_0 \qquad (3)$$

Table 1 summarizes the possible values of the conductance.

| The value of J | $G_f$, ferromagnetic interaction | $G_a$, antiferromagnetic interaction |
|---|---|---|
| J=1/2 | $G_f = 3/4 G_0$ | $G_a = 1/4 G_0$ |
| J=1 | $G_f = 2/3 G_0$ | $G_a = 1/3 G_0$ |
| J=3/2 | $G_f = 5/8 G_0$ | $G_a = 3/8 G_0$ |
| J=2 | $G_f = 3/5 G_0$ | $G_a = 2/5 G_0$ |
| J=5/2 | $G_f = 7/12 G_0$ | $G_a = 5/12 G_0$ |
| $J = \infty$ | $G_f = 1/2 G_0$ | $G_a = 1/2 G_0$ |

In the case of the ferromagnetic interaction, which is likely to be realized in the experiment, the height of the sub-step is seen to decrease with the number of the impaired electrons localized on the contact and to reach the value of $1/2G_0$ if this number becomes infinite (in fact, it is almost $1/2G_0$ for $J=3$). Besides, the number of the unpaired electrons is expected to be dependent on the length of the QPC, being small for short and large for long contacts, respectively. Thus, for the short contacts the conductance should be about $0.7G_0$ in accordance with the experimental data, while for the long wires it should attain the value of $0.5G_0$. This result corresponds perfectly to the experimental observations by Reilly et al [8]. The application of the external magnetic field leads to the spin polarization of both propagating and localized carriers thus transforming the conductance into $G = e^2/h$ for all values of $J$.

Now the conductance can be calculated in more rigorous way. We perform our analysis within frameworks of the following model. The movement of the propagating electrons is supposed to be purely one-dimensional, i.e. the diameter of the quantum D wire is taken to be small enough to provide the validity of the following condition

$$\frac{mD^2 E_F}{\pi^2 \hbar^2} < 1 \tag{4}$$

where the Fermi energy of carriers can be estimated as $E_F = \frac{\pi \hbar^2}{m}\left(n_{2D} - \frac{\pi}{2D^2}\right)$. Besides, the temperature should be small to prevent the thermal mixing of the subbands, $\frac{mD^2 kT}{\pi^2 \hbar^2} \ll 1$. The conductance of the system appears to be calculated by means of the Landauer-Buttiker formula where the transmission coefficients depend on the Fermi energy and the spin of carriers.

If the external magnetic field is absent and electrons in the ingoing and outgoing leads are unpolarized, the density matrix of the system containing the free propagating electrons, and the localized spin before their interaction reads

$$\boldsymbol{\rho}_{in} = \boldsymbol{\rho}_e \otimes \boldsymbol{\rho}_J = \frac{1}{2}\left(|\uparrow_e\rangle\langle\uparrow_e| + |\downarrow_e\rangle\langle\downarrow_e|\right) \otimes \left(\frac{1}{2J+1}\sum_{m=0}^{2J}|J-m\rangle\langle J-m|\right) \tag{5}$$

Thus, there are 4J+2 possible mutual orientations of the spin of the propagating and localized electrons. For each of them after passing of the region of the QPC, the spin of the propagating electron can be either conserved or inversed due to the exchange interaction. For example, the situation when the propagating electron initially has spin projection *-1/2* on the arbitrary selected axis, while the spin projection of the localized spin on the same axis is $J-m+1$, can be imagined. After the interaction the spin projections can rest the same, $-1/2$ and $J-m+1$, or spin-flip can occur, and spin projections will be $+1/2$ and $J-m$. The conductance at zero temperature can be thus calculated as

$$G_{T=0}(E) = \frac{e^2}{4h(2J+1)} \sum_{m=0}^{2J} \left[ \left| A_{\left(-\frac{1}{2};J-m+1\right) \to \left(-\frac{1}{2};J-m+1\right)} \right|^2 + \left| A_{\left(-\frac{1}{2};J-m+1\right) \to \left(\frac{1}{2};J-m\right)} \right|^2 + \left| A_{\left(\frac{1}{2};J-m\right) \to \left(\frac{1}{2};J-m\right)} \right|^2 + \left| A_{\left(\frac{1}{2};J-m\right) \to \left(-\frac{1}{2};J-m+1\right)} \right|^2 \right]$$

(6)

where A denotes the transmission amplitudes dependent on the Fermi energy of carriers. The indices of the transmission amplitudes denote the spin state of the propagating and localized electrons before and after interaction. Thus, the amplitudes $A_{\left(-\frac{1}{2};J-m+1\right) \to \left(-\frac{1}{2};J-m+1\right)}$ and $A_{\left(\frac{1}{2};J-m\right) \to \left(\frac{1}{2};J-m\right)}$ describe the spin-conservative passing of the carrier through QPC, while $A_{\left(-\frac{1}{2};J-m+1\right) \to \left(\frac{1}{2};J-m\right)}$ and $A_{\left(\frac{1}{2};J-m\right) \to \left(-\frac{1}{2};J-m+1\right)}$ correspond to the passing accompanied by a spin-flip. Formula (6) can be easily generalized for the case of the nonzero temperature

$$G(T,\mu) = \int_0^\infty G_{T=0}(E) \left( -\frac{\partial f(T,E,\mu)}{\partial E} \right) dE \qquad (7)$$

Where $\mu$ denotes the chemical potential, $f(T,E,\mu)$ is the Fermi distribution.

To determine the values of the transmission amplitudes in Eq(6), it is necessary to specify the Hamiltonian of the interaction between the propagating carrier and localized spin $J$. In the present work we suppose that they interact only in the region of the length $L$ (dimension of QPC), whereas in the other space the interaction is taken to be absent. The model Hamiltonian can be thus represented in the following form

$$H = \begin{cases} \dfrac{\hbar^2 k^2}{2m}, & x<0, x>L \\ \dfrac{\hbar^2 k^2}{2m} + V_{dir} + V_{ex}\boldsymbol{\sigma}\cdot\mathbf{J}, & x \in [0,L] \end{cases} \qquad (8)$$

For the ferromagnetic interaction, $V_{ex}>0$, while for the antiferromagnetic interaction, $V_{ex}<0$. To calculate the spin-dependent transmission amplitudes, the Hamiltonian in the region of the QPC is followed to present in the matrix form, using the basis of the $4J+2$ vectors

$$|\psi_{2m}\rangle = \left|\frac{1}{2}; J\right\rangle; \ldots |\psi_{2m}\rangle = \left|-\frac{1}{2}; J-(m-1)\right\rangle; \quad |\psi_{2m+1}\rangle = \left|\frac{1}{2}; J-m\right\rangle, \ldots |\psi_{4J+2}\rangle = \left|-\frac{1}{2}; -J\right\rangle \quad (9)$$

where $m = 1, \ldots, 2J$. The matrix of the Hamiltonian is of importance to represent in a block-diagonal form because of the spin conservation

$$H_{lk} = V_l^{(1)} \delta_{lk} + V_l^{(2)} \left(\delta_{l,k+1} + \delta_{l+1;k}\right) \quad (10)$$

Where the parameters $V_l^{(1,2)}$ read

$$V_1^{(1)} = V_{4J+2}^{(1)} = \frac{\hbar^2 k^2}{2m} + V_{dir} + V_{ex} J; \quad (11a)$$

$$V_1^{(2)} = V_{4N+2}^{(2)} = 0 \quad (11b)$$

$$V_{2m}^{(1)} = \frac{\hbar^2 k^2}{2m} + V_{dir} - V_{ex}(J - m + 1) \quad (11c)$$

$$V_{2m+1}^{(1)} = \frac{\hbar^2 k^2}{2m} + V_{dir} + V_{ex}(J - m) \quad (11d)$$

$$V_{2m}^{(2)} = V_{ex}\sqrt{m(2J - m + 2)} \quad (11^e)$$

$$V_{2m+1}^{(2)} = 0 \quad (11f)$$

This Hamiltonian can be reduced to the diagonal form $H_{lk} = \varepsilon_l \delta_{lk}$, where

$$\varepsilon_1 = \varepsilon_{4J+2} = \varepsilon_{2m+1} = \frac{\hbar^2 k^2}{2m} + V_{dir} + V_{ex} J \quad (12a)$$

$$\varepsilon_{2m} = \frac{\hbar^2 k^2}{2m} + V_{dir} - V_{ex}(J + 1) \quad (12b)$$

The value given by the formula (12a) corresponds to the total spin of the localized plus propagating electron that is equal to $S_1 = J + 1/2$, while the value given by (12b) corresponds to the total spin $S_2 = J - 1/2$.

To determine the transmission amplitudes, it is also necessary to obtain the general expression for the wavefunction for all possible mutual orientations of the spin of the propagating electron and the localized spin. For example, the combination of the spin projections that consists of the spin projection of the electron falling to QPC equal to *-1/2* and the projection of the localized spin corresponding to *J-m+1* is more of interest. In this case the wavefunction reads

$$\Psi_1(x) = \begin{pmatrix} 0 \\ 1 \end{pmatrix} e^{ik_F x} + B_{\left(-\frac{1}{2};J-m+1\right)\to\left(-\frac{1}{2};J-m+1\right)} \begin{pmatrix} 0 \\ 1 \end{pmatrix} e^{-ikx} + B_{\left(-\frac{1}{2};J-m+1\right)\to\left(\frac{1}{2};J-m\right)} \begin{pmatrix} 1 \\ 0 \end{pmatrix} e^{-ik_F x}, \quad x<0$$

$$\Psi_2(x) = \mathbf{X}_m^{(1)}\left(C_{+1m}e^{ik_1 x} + C_{-1m}e^{-ik_1 x}\right) + \mathbf{X}_m^{(2)}\left(C_{+2m}e^{ik_2 x} + C_{-2m}e^{-ik_2 x}\right), \quad x\in[0;L] \qquad (13)$$

$$\Psi_3(x) = A_{\left(-\frac{1}{2};J-m+1\right)\to\left(-\frac{1}{2};J-m+1\right)} \begin{pmatrix} 0 \\ 1 \end{pmatrix} e^{ik_F x} + A_{\left(-\frac{1}{2};J-m+1\right)\to\left(\frac{1}{2};J-m\right)} \begin{pmatrix} 1 \\ 0 \end{pmatrix} e^{ik_F x}, \quad x>L$$

where $A$ and $B$ are transmission and reflection amplitudes respectively, $k_F$ is the Fermi wavenumber of the carrier inside the leads, the wavenumbers $k_1$ and $k_2$ correspond to the eigen energies of the carrier in the region where exchange interaction undergoes (see formulae 12a, 12b)

$$k_1 = \sqrt{\frac{2m}{\hbar^2}\left[E_F - V_{dir} - V_{ex}(J+1)\right]} \qquad (14a)$$

$$k_2 = \sqrt{\frac{2m}{\hbar^2}\left[E_F - V_{dir} + V_{ex}J\right]} \qquad (14b)$$

$\mathbf{X}_m^{(1,2)}$ are eigenvectors of the m-th block of the Hamiltonian and read

$$\mathbf{X}_1^{(m)} = \frac{1}{\sqrt{2J+1}}\begin{pmatrix} \sqrt{2J-m+1} \\ -\sqrt{m} \end{pmatrix}; \quad \mathbf{X}_2^{(m)} = \frac{1}{\sqrt{2J+1}}\begin{pmatrix} \sqrt{m} \\ \sqrt{2J-m+1} \end{pmatrix} \qquad (15)$$

All other possible mutual orientations can be also considered.

The expressions (12)- (14) together with the continuity condition of the wavefunction and its derivative in the points x=0 and x=L allow the determination of the transmission amplitudes of the spin conservative process $A_{\left(-\frac{1}{2};J-m+1\right)\to\left(-\frac{1}{2};J-m+1\right)}$ and of the spin- flip process $A_{\left(-\frac{1}{2};J-m+1\right)\to\left(\frac{1}{2};J-m\right)}$. Then, the conductance can be calculated using the formula (6).

Figs 1a and 1b show the dependence of the conductance of the QPC on the chemical potential for different values of the $J$. The parameters of the calculation were taken as follows: the effective mass of the carrier m=0.06$m_e$, the temperature T=10 K, the length of the contact $L = L_0 J$ with $L_0$ = 10 nm (This means that the length of the contact is proportional to the number of the electrons localized on it). The values of the exchange and direct matrix elements were estimated as $V_{ex} \approx \mp e^2/L$, $V_{dir} = \mp 1,1\cdot V_{ex}J$. Figs 1a and 1b show the conductance for the ferromagnetic and the antiferromagnetic interaction, respectively. The formation of the additional plateau, the height of which corresponds to one given by the formulae 2 and 3, are clearly seen.

In conclusion, we have analyzed the conductance of the QPC containing a large localized spin. The additional plateau of the conductance has been shown to be formed in this system. The conductance value of this plateau appeared to depend on the total spin of the contact and to decrease from $3e^2/2h$ to $e^2/h$ with increasing of *J*, which is in a good agreement with experimental observations.

This work has been supported by SNSF in frameworks of the programme "Scientific Cooperation between Eastern Europe and Switzerland, Grant IB7320-110970/1.

**Captions**

**Fig. 1.**

The first step of the quantum conductance staircase vs the chemical potential of carriers that reveals the fractional variations of the "0.7 ($2e^2/h$)" feature as a function of the localized spin, *J*, inside the quantum point contact, QPC, in the case of the ferromagnetic (a) and the antiferromagnetic (b) exchange interaction.

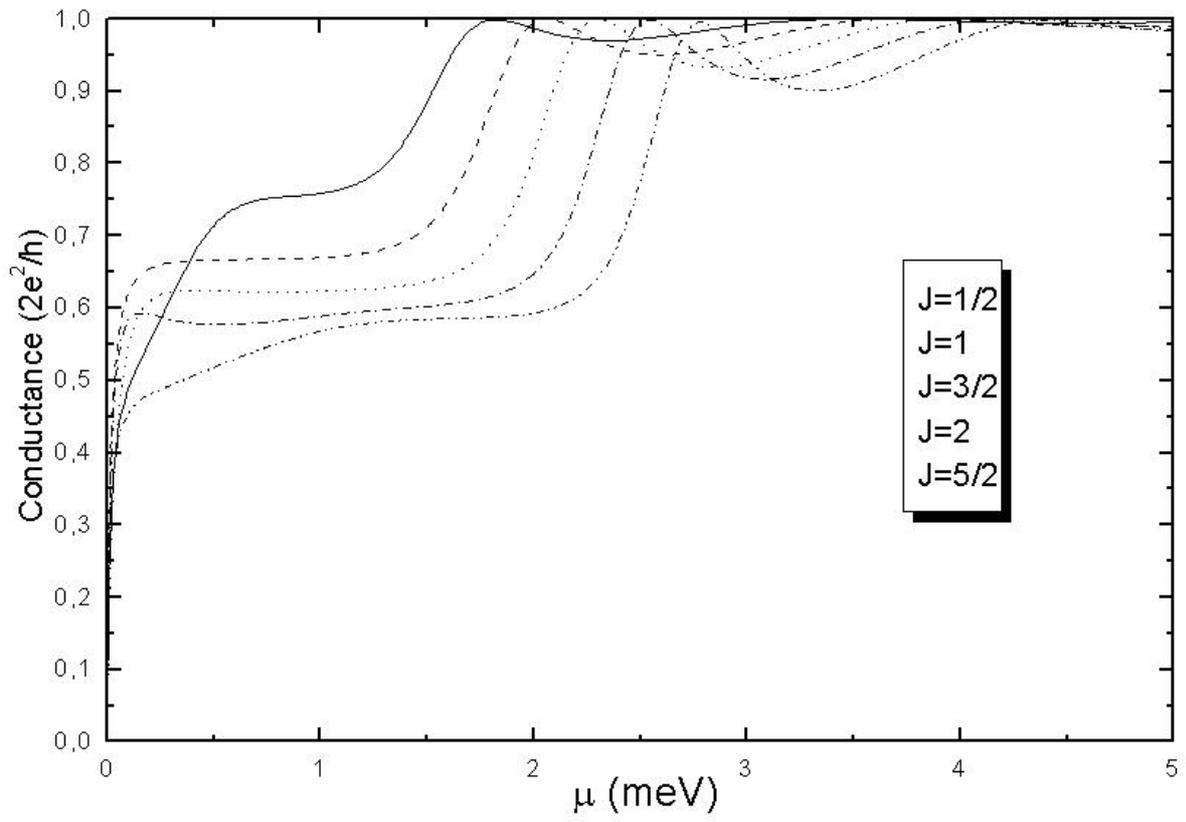

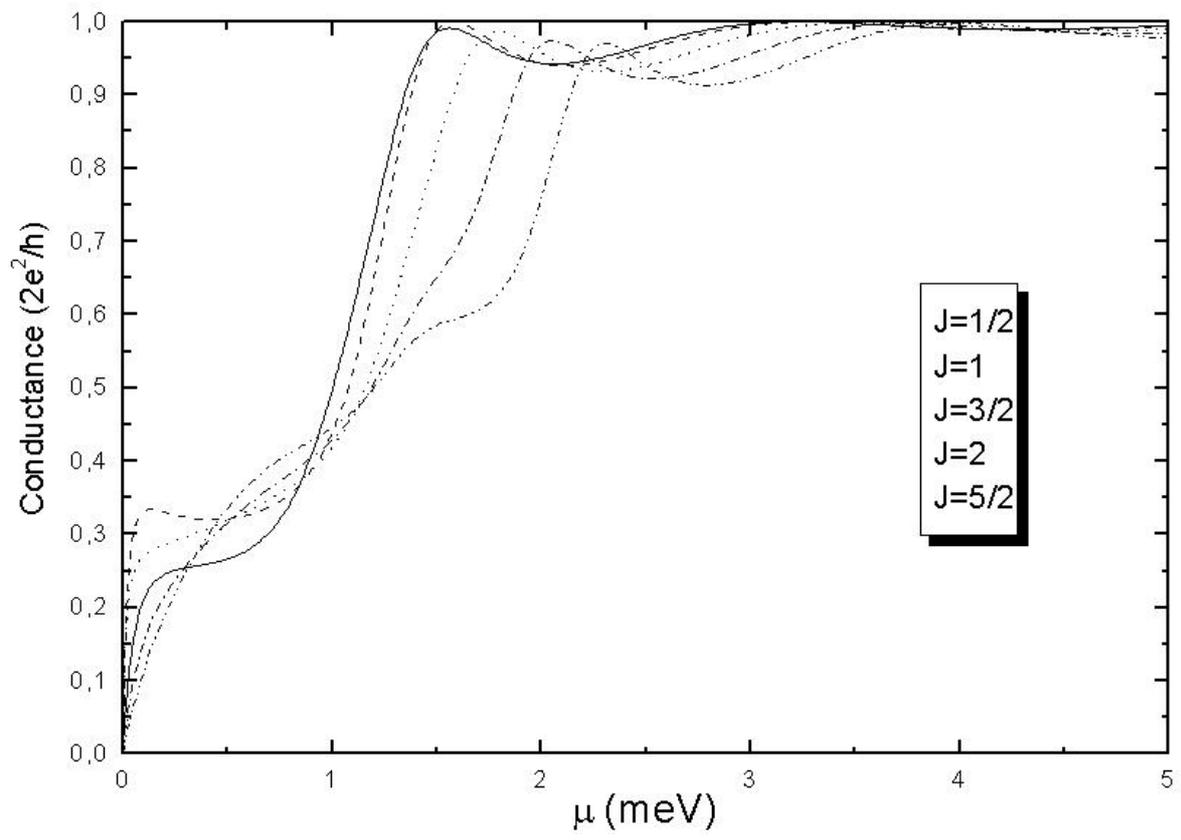